# Optical Tests of Nanoengineered Liquid Mirrors

**Hélène Yockell-Lelièvre[2], Ermanno F. Borra[1], Anna Ritcey[2],**

**Lande Vieira da Silva Jr[1].**

Centre d'Optique, Photonique et Lasers, Université Laval, Québec, Qc, Canada G1K 7P4.

Their email adresses are: borra@phy.ulaval.ca, Anna.Ritcey@chm.ulaval.ca,

lvidasil@phy.ulaval.ca, hyockell@chm.ulaval.ca

(SEND ALL EDITORIAL CORRESPONDANCE TO: E.F. BORRA)

OCSIS: 000.3110, 240.6700, 260.3910, 350.1260

1 Département de Physique, Genie Physique et Optique

2 Département de Chimie.

Abstract

We describe a new technology for the fabrication of inexpensive high-quality mirrors. We begin by chemically producing a large number of metallic nanoparticles coated with organic ligands. The particles are then spread on a liquid substrate where they self-assemble to give optical quality reflective surfaces. Since liquid surfaces can be modified with a variety of means (e.g. rotation, electromagnetic fields), this opens the possibility of making a new class of versatile and inexpensive optical elements that can have complex shapes which can be modified within short time scales. Interferometric measurements show optical quality surfaces. We have obtained reflectivity curves that show 80% peak reflectivities.We are confident that we can improve the reflectivity curves, for theoretical models predict higher values. We expect that nanoengineered liquid mirrors should be useful for scientific and engineering applications. The technology is interesting for large optics, such as large rotating parabolic mirrors, because of its low cost. Furthermore, because the surfaces of ferrofluids can be shaped with magnetic fields, one can generate complex, time varying surfaces difficult to make with conventional techniques.

# 1. Introduction

Liquid surfaces follow equipotential surfaces so that a reflective liquid makes a mirror that has the shape of the equipotentials. Liquid mirrors therefore offer new possibilities for the fabrication of unique optical elements. For example, the fact that the surface of a liquid rotating in a gravitational field takes the shape of a parabola has been used to make large inexpensive parabolic mirrors. For a given diameter, the cost of a diffraction limited mercury liquid mirror is almost two orders of magnitudes lower than the cost of a glass mirror, making it very competitive for some applications. The technology is young but its quality and robustness have been documented in the laboratory[1,2] as well as in observatory settings[3,4,5] . However, liquid metals have several inconveniences (e.g. high-density, toxicity, etc..) so that it would be advantageous to replace them with other liquids having better characteristics. Furthermore, liquid surfaces can be shaped with other means than rotation. For example, the surfaces of ferrofluids can be shaped with magnetic fields [6]. Unfortunately, mercury and some gallium alloys [7] are the only metals that are liquid at room temperature so that we have a limited range of parameters (e.g. magnetic susceptibilities) to work with if restricted to liquid metals.

To extend the range of useful liquids and the range of available parameters, such as magnetic susceptibility, we have begun a research program that endeavors to find techniques that allow the deposition of high-reflectivity metallic surfaces on liquids. In a recent paper[8] we have discussed the characteristics of Metal Liquid Like Films (MELLFs) trapped at the interface between two liquids. MELLFs were originally developed by chemists to study the characteristics of the chemicals that coat the metallic grains. Realizing their potential use in optical applications, we have started a research program to investigate their optical properties and refine the chemical techniques for optical applications.

In this article, we describe the basic technology (section 2) and discuss the optical characteristics of MELLFs deposited at the top surface of a liquid (section 3). The spreading of the reflecting layer at the top surface represents a significant improvement as compared to our previous mirrors in which the nanoparticles were assembled at the interface between two liquids. In the present configuration, the reflective nanoparticle coating is the first surface encountered by the incoming electromagnetic wave.

## 2. Preparation of silver particles and deposition of reflective films on liquids

Stable interfacial suspensions of silver nanoparticles have been described in the literature and are frequently referred to as Metal Liquid-Like Films, or MELLFs[9,10]. These unique systems combine the optical properties of metals with the fluidity of a liquid suspension and are therefore well adapted to applications in the field of liquid optics. The fabrication of a MELLF involves the creation of silver nanoparticles, generally by chemical reduction of a silver salt in aqueous solution, and the subsequent coating of the particles with a strong metal-bonding organic molecule, a ligand. When coated, the particles are no longer stable in the aqueous phase and spontaneously assemble at the water-organic interface.

The MELLFs investigated in the present study were prepared in the following way: Silver nanoparticles were first prepared by the chemical reduction by citrate of silver nitrate in aqueous solution following a known procedure[11]. Two different preparations were considered: in one case the temperature was maintained precisely at $100^{o}$C, whereas in the other the temperature was less well controlled and slightly lower (around $95^{o}$C). The silver nitrate solution was brought to a boil ($100\,^{o}$C) before the addition of the reducing agent. In the case of more precise temperature control, the citrate solution was also heated to boiling before its addition, thus permitting the temperature to be maintained at $100^{o}$C throughtout the reduction process.  In the second preparation method, a room temperature citrate solution was added to the silver nitrate solution, resulting in a temporary drop in the temperature. The resulting aqueous suspension of metallic particles was then vigorously shaken for 5 minutes together with a solution of either 1,10-phenanthroline, 2,9-dimethyl-1,10-phenanthroline or 2,2-dipyridyl in 1,2-dichloroethane. This step provokes the spontaneous formation of a concentrated fluid phase of particles at the interface between the two liquids.  Finally, the concentrated interfacial suspension of particles was collected by removal of the excess aqueous and organic phases, washed  and poured onto a water surface.  At this point, the residual 1,2-dichloroathane in the MELLF evaporates, leaving a thin dried film of ligand-coated nanoparticles.  Although this film is not truly liquid-like, it is able to withstand small deformations without breaking.

It must be noted that, although the work reported below refers only to mirrors spread on pure water, certain characteristics of the liquid support (e. g. viscosity or magnetic susceptibility) can be easily varied by the introduction of suitable additives in the water (e.g. water-soluble polymers or colloidal particles). More complete details of the chemical preparation and characterization of the MELLFs will appear elsewhere.

The shape and size of the particles obtained from the two different preparations  are shown in the transmission electron microscope (TEM) images in Fig. 1. Keeping a constant temperature of 100 degrees Celsius predominantly produces spherical particles having variable diameters of a few 10 nm. If we reduce the temperature by a few degrees during reduction, we produce elongated particles of the order of 100 nm.  Obviously, we have some measure of control of the size and shapes of the particles by controlling the temperature during the first few minutes of the chemical reduction  that produces them .

## 3. Optical tests of nanoengineered mirrors

## 3.1 Interferometry

The Laval liquid mirror laboratory is presently carrying out work to develop large rotating parabolic liquid mirrors that use metal coated liquids as well as magnetically deformable adaptive mirrors made of coated ferrofluids [6]. Most of this work is in progress and will be reported elsewhere at later times. Because the aim of the present article is to introduce the basic concept and coating technology, we shall only discuss flat water-based mirrors. We have made and tested several flat liquid mirrors. We have carried out interferometric tests at a wavelength of 632.8 nm with a Fizeau interferometer made by the Zygo corporation (model GPI-XP).

Fig. 2a shows an interferogram obtained by measuring a 6.5 cm diameter sample of one of our dried MELLF surfaces. The lateral resolution on the surface of the liquid was 0.1 cm. The fringes can be compared to those obtained, in the same conditions, by measuring a $\lambda/20$ glass mirror (Fig. 2b). We can readily see that the surfaces have comparable qualities. This is to be expected since the surface of a liquid follows an equipotential surface to a very high precision, as shown by the work on mercury liquid mirrors[1]. On the other hand, there are particular problems that can arise with the coating technology. For example, large surfaces can be marred by a few bubbles, as well as a few scar-like defects caused by evaporating water drops that leave solid residues. The fractions of the surfaces affected are not large ( of the order of a few percents) and we are presently developing techniques to minimize these defects.

Obviously, interferometry does not provide information about the surface roughness over distances smaller than the resolution of the interferometer. Scattered light measurements will be made to quantify the scattering, but results are not available at this time. Atomic Force Microscope analyses of samples deposited on glass indicate root mean square (RMS) surface deviations of the order of 20 nm. Although glass mirrors can have smoother surfaces, this value corresponds to about $\lambda/20$ at 500 nm and predicts little scattered light. On the other hand, some scattered light can also be caused by clumps of coalesced particles. As discussed below, our reflectivity and transmission curves show that this is particularly a problem if we use a small concentration of the protective ligands. The curves give some indications of the importance of scattered light since the lost energy is due to a combination of scattered light and absorbed light.

## 3.2 Reflectance and transmittance curves

We measured the reflectivity and transmission curves of several of our samples as a function of wavelength with a spectrophotometer (Varian model Cary 500 Scan). The reflectivity curves were normalized by comparing them to the curves obtained on a sample of clean mercury. While the reflectivity curves shown are measured on the reflective liquid, the transmission curves are measured on samples deposited on a glass microscope slide. The sample is transferred from the water surface simply by sliding the glass under the surface and then lifting it horizontally to remove a section of the MELLF. A freshly deposited dried MELLF rests on top of the thin layer of water that wets the slide. This layer evaporates within a few minutes, leaving a dried film on the glass. Transmission measurements were carried out on both water-deposited and glass-deposited samples, and no significant differences were found. The reflectivity curves for a typical dried MELLFs is shown in Fig. 3 (curve a). Maximum reflectivities of the order of 80% are obtained near 700 nm. Lower reflectivity is observed at both higher and lower wavelengths. The reflectivity losses can be attributed to dissipative absorption, scattering and the fact that the reflective layer is relatively thin and transmits some of the light. We have used the transmission curves to evaluate the percentage of light lost to the different factors. The difference between the reflectivity and transmission curves is plotted in Fig. 3 (curve b) indicates that transmission alone is not responsible for the loss. There obviously is, in addition, a combination of scattering and dissipative absorption. At shorter wavelengths, important losses due to the surface plasmon absorption near 320 nm are evident. above 700 nm transmission increases with increasing wavelength, varying from about 10% at 500nm to 30% at 1200 nm.

The reflectivity curves give information on the physical characteristics of the reflecting layers and useful indications of what is needed to improve the optical properties of the films. For example, Fig. 4 illustrates the influence of the size and shape of the particles, showing results for samples prepared from the two different preparations of colloidal particles, formed by reduction at either boiling or slightly lower temperature. Though they are quantitatively different the reflectivity curves exhibit similar tendencies. Both show that the lowest reflectivity value is in the plasmon frequency for bulk silver at ~320 nm. In the visible region of the spectrum there is a peak in the reflectivity at ~500 nm followed by a slight decrease which gives it a hunchback profile. The values then increase to their highest value ~800 nm to decrease monotonically with increasing wavelength. These are the typical features we see in the spectra of all of our MELLFs.

The three ligands that we have tested (1,10-phenanthroline, 2,9-dimethyl-1,10-phenanthroline and 2,2-dipyridyl ) had a relatively small effect on the form of the reflectivity curves. However, considering the limited number of ligands tested, and their obvious structural similarities, we cannot make definitive statements regarding the importance of the chemical composition of the ligands. We are currently investigating a variety of additional ligands.

We find that reflectivity decreases if we lower the concentration of ligands. Presumably this increases the amount of scattered light since too little ligand may leave unprotected areas on the silver grains that clump together to give large irregularly-shaped aggregates that scatter light.

## 3.3 Model of reflectance properties

To understand the shape of our reflectivity curves and to try and find ways to improve them, we use the theoretical framework described by [12]. They developed a model, which they called the TKF approximation, to describe the optical behavior of composite systems made of particles embedded in a dielectric. It corrects the bulk optical constants of a metal for the particle size[13] and for the volume fraction (filling factor) of the particles in the matrix[14]. It uses the corrected values with the Fresnel equations to calculate the reflectivity of the system. Whenever the particle size is comparable to or smaller than that of the mean free path of the conduction electrons in the bulk material ($l_{inf}$), we need to correct the dielectric constants of the system. This must be so because the collisions of these electrons with the particle surfaces lower the effective mean free path leading to an increase in the collision frequency. This is the classical approach. The new dielectric constant is then a function of the incident light frequency and of the radius of the particle [13].

Let $\tau = 3.7 \times 10^{-14}$ seconds be the collision time (relaxation time) for bulk silver at 273 K . The mean free path of the conduction electrons in the bulk material is then given by $l_{inf} = v_f \, \tau$ . One obtains $l_{inf} = 52$ nm with the fermi velocity $v_f = 1.4 \times 10^6$ m/s .

The value for $l_{inf}$ is different if the collision bulk frequency, which is given by $\omega_o = (\tau)^{-1}$, changes. If particle sizes are comparable to or smaller than $l_{inf}$ the collision time $\tau$ changes. The new collision time is a function of the radius of the particle[13] and is given by $\tau_{corr} = R / v_f$. Finally, if the collision time changes, so does the collision frequency which is then corrected for the particle size [13] to give $\omega_c(R) = \omega_o + (\tau_{corr})^{-1}$ and $\omega_c(R) = \omega_o + (v_f / R)$. The change in the collision frequency affects the optical constants. For silver, the real part of the optical constant (n) is inversely proportional to the radius of the particle. The imaginary part (k) remains basically the same except for the plasmon region of the spectra. The TKF model predicts that reflectivity will decrease for smaller particle radius in the 400 – 600 nm region of the spectra. Changes in the particle size do not change the reflectivity from 700 nm on. Further discussion is given in reference [13]. Note however that our experimental curves in Fig. 2 disagree with this last statement. On the other hand, the disagreement may partially be due to differences between the model and our samples, as discussed in the next section.

The correction of the optical constants for variations in the filling factor , defined as the volume ratio of the sum of the metallic particles to the total volume of the matrix and particles, is taken into account using the expressions given by Torquato [14]. He considers multipolar electromagnetic interactions between spherical particles, which increase with the high filling factor: the higher the filling factor, the higher the electrical conductivity and the higher the reflectivity. This can be understood by the increase of clusters of particles[15] resulting on an increased electrical conductivity.

Unfortunately, there is an upper limit to the filling factor; partly because the particles are wrapped up in a thin layer of ligand (about 1 nm thick) and partly because of the geometrical limit to the packing of particles that are not orthorombic (e.g. boxes). The filling factor therefore depends on the shapes and distributions of the particles. For example, the maximum filling factor value for a random close packing of spheres that have the same diameters would be ~ 63% [16]. It may be possible to increase the filling factor by mixing spheres of different sizes. Torquato [14] gives a three point parameter that varies with the filling factor for each different spatial arrangement like simple cubic, face centered cubic among others, allowing us to consider these different packings in the TKF model calculations.

## 3.4 Comparison of reflectivity data to models

Reflectivity curves computed with the TKF model show the effect of the sizes of the particles (assuming spheres) as a function of diameters for a filling factor of 55% in Fig. 5. Theory predicts that reflectivity increases from 1nm to 30 nm diameters, while there is little size effect for spherical particles larger than 30 nm. However, we see that the experimental reflectivity curves in Fig. 4 are significantly affected by the sizes and shapes of the particles. The discrepancy may be due to inadequacies of the TKF theory or to the fact that the theory assumes smooth spherical particles all having the same radii, while the electron microscope images show that the particles are not spherical and have different sizes (Fig. 1). Fig. 6 shows reflectivity curves for particles having diameters of 20 nm and three different filling factors. We used Torquato's[14] values for randomly distributed spheres. Figure 5 shows that reflectivities longward of 650 nm can be greatly improved by increasing the filling factor. It must be noted that hard spheres having the same diameter cannot give filling factors higher than 74%, so that the 95% curve is misleading for such a situation. On the other hand, a filling factor larger than 75% may be obtained by using a distribution of diameters, since the smaller spheres will fill the regions in between the larger spheres.

Notwithstanding the disagreements mentioned in the preceding discussion, the TKF model points the way to improve the reflectivity of our liquids. Fig. 6 shows that the major improvements will come by increasing the filling factor.

## 4. Conclusions.

This article uses a novel chemical technique that coats the air-water interface with a reflective layer of self-assembling metallic nanoparticles. We have succeeded in coating the top surface of water with a layer of silver grains coated with a ligand. This is a significant improvement over results presented in earlier work[8] which trapped the reflecting layer at the interface between two liquids since now the reflecting layer coats the first surface seen by light. Interferometry shows that the coated surfaces have good optical qualities. Reflectivity measurements show peak values of the order of 80%, comparable to the reflectivity of mercury (about 80%). Reflectivity is lower than the reflectivity of silver or aluminum coated glass but one must consider that, in practice, coated glass mirrors are seldom recoated, hence degrade with time as they get covered with dust and dirt. This is particularly true for telescope mirrors used in astronomy, the space and atmospheric sciences which are exposed unprotected to the elements. On the other hand, our reflecting layers are easy to generate and inexpensive so that they could be refreshed often. The 80% peak reflectivity was obtained effortlessly and we are confident that reflectivities can be improved with additional work currently in progress.

Experimental and theoretical curves based on the TKF model are in qualitative agreement but disagree in the details. We can probably "hand-wave" the differences by invoking "real-world" differences between our results and the TKF calculations. The TKF model assumes smooth spherical particles having the same diameters. In our case, the electron microscope images (Fig 1) show that the particles can be elongated, are not smooth, and have different radii. Furthermore, our particles tend to clump together, so that we do not have any of the idealized distributions assumed in the computations. Finally there is a loss due to scattering by large aggregates of particles that we cannot evaluate at this time. Still, the models should be useful to show the ways to increase reflectivity. They suggest that we should increase the filling factor to improve reflectivity.

We expect that nanoengineered liquid mirrors should be useful for scientific and engineering applications. The technology is interesting for large optics, such as large parabolic mirrors, because of its low cost. Consider that the mirror basically consists of a 1-mm thick layer of liquid coated with a 150 nm thick layer of nanoparticles. A 10-m diameter mirror would thus weigh of the order of 100 kg. Because the cost of the mirror and supporting hardware (e.g. container and bearing) depends on the weight of the mirror, it should be quite inexpensive (of the order of a few tens of thousands of dollars). The costs of the chemical involved are negligible and the process not overly demanding of human time. The cost of an astronomical observatory would then be totally dominated by the cost of the building and the instrumentation. For all essential purposes one would get the mirror for free. Note also that, while mercury mirrors cannot be tilted, thus limiting their usefulness, rotating mirrors that use viscous liquids can be tilted[8] . Coated ferrofluids[6] are especially interesting for practical applications because they promise to make versatile low-cost mirrors. Because the surfaces of these mirrors can be shaped with magnetic fields, one can generate complex, time varying surfaces difficult to make with conventional techniques.

     This is a very young technology and there is a significant quantity of work to be done before it reaches the performance and robustness expected for practical applications. We are presently working to improve it. Recently, we have developed a technique to coat the surface of oils with MELLFs, to be reported elsewhere. We are also investigating the use of other metals (e.g. gold and aluminum). The work carried out so far has been relatively simple and we have reached several milestones with relative ease, boding well for the future.

**AKNOWLEDGEMENTS.**

This research has been supported by grants to E. Borra and A. Ritcey from the Natural Sciences and Engineering Research Council of Canada. and The Canadian Institute for Photonics Innovation. Lande Vieira da Silva Junior is financially supported by a scholarship from the ' Conselho Brasileiro de Desenvolvimento Científico e Tecnológico', CNPq – Brasil.

**FIGURE CAPTIONS.**

Figure 1 : Transmission electron microscopy (TEM) images of MELLF samples prepared from a silver colloid reduced at constant boiling temperature ($100^{o}$C, a) and slightly lower ($95^{o}$C, b) temperatures.

Figure  2 : Interferograms of a clean optically flat mirror (a) and of a 6.5 cm diameter MELLF sample dried at the water-air interface (b). We can readily see that the surfaces have comparable qualities.

Figure 3 : Reflected light (a) and "lost" light (100 - Reflected - Transmitted), corresponding to scattered and absorbed light (b), measured on a dried MELLF sample prepared from a colloid reduced at $100^{o}$C.

Figure 4: Reflected light as a function of wavelength measured on dried MELLF samples prepared from silver colloids reduced at $100^{o}$C (1) and $95^{o}$C (2).

Figure 5: Theoretical reflectivity curves calculated with the TKF model. They show the effect of the sizes of spherical particles (assuming spheres) for a filling factor of 55%.

Figure 6: Reflectivity curves calculated with the TKF model corresponding to filling factors of 55%, 75% and 95% and assuming 20-nm radius spherical silver particles. We used Torquato's values for randomly distributed spheres.

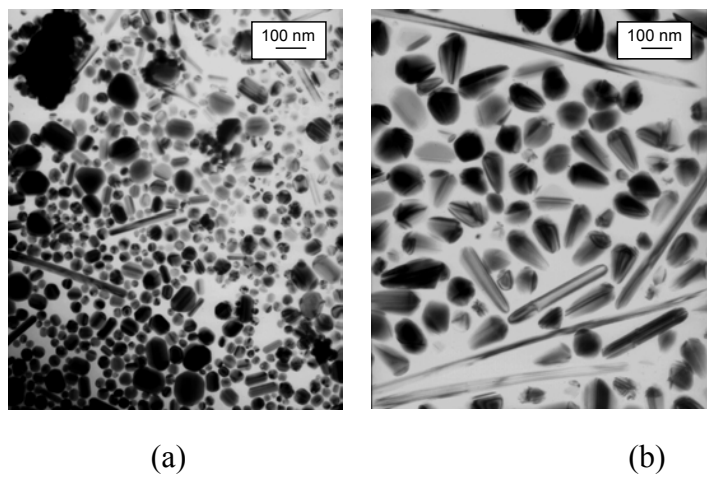

(a)                              (b)

FIGURE 1

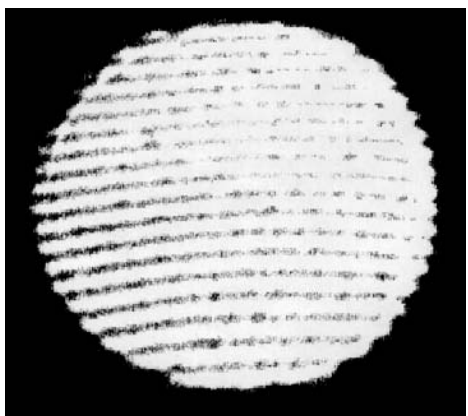
a)

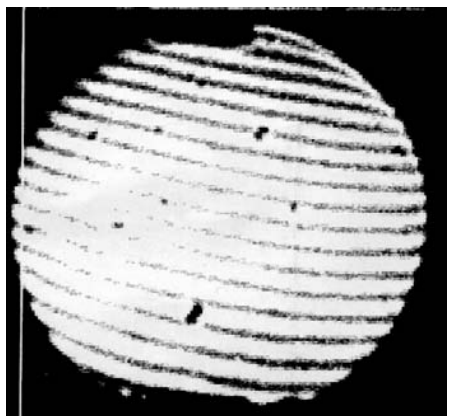
b)

FIGURE 2

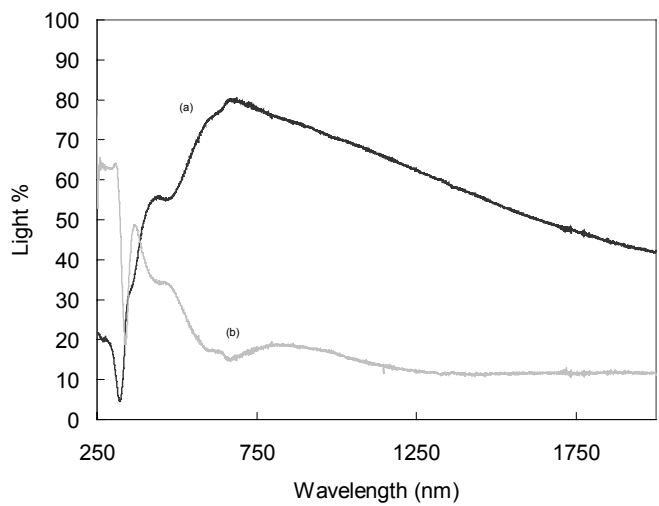

FIGURE 3

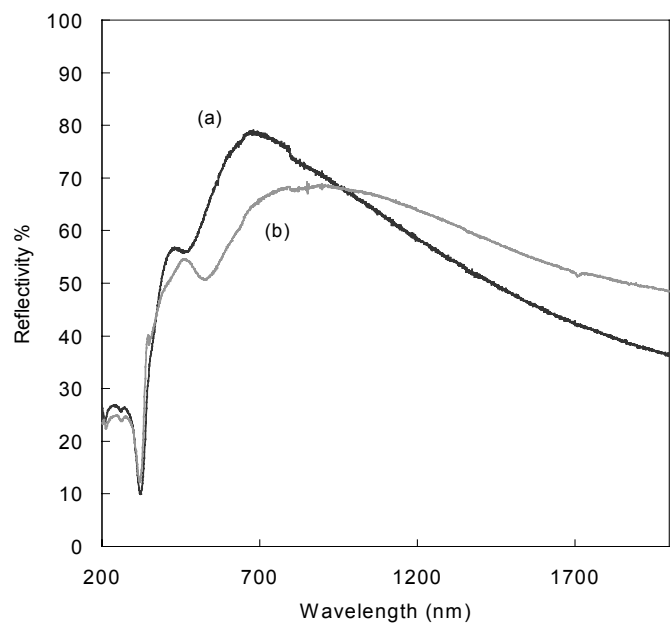

FIGURE 4

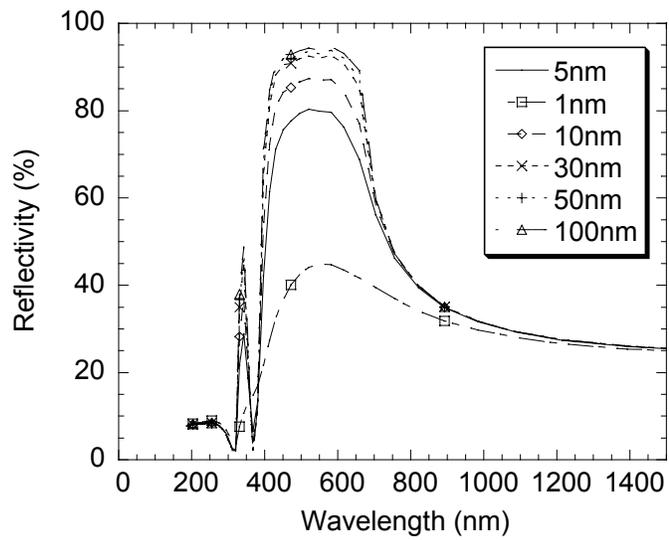

FIGURE 5

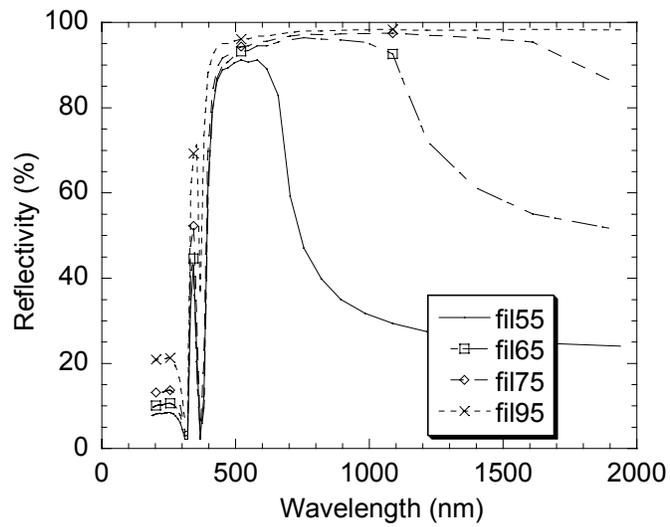

FIGURE 6